\documentclass[runningheads]{llncs}
\usepackage[T1]{fontenc}
\usepackage{graphicx,verbatim}
\usepackage{multirow}
\usepackage{makecell}
\usepackage{xcolor}
\usepackage{colortbl}
\usepackage{booktabs}
\usepackage{hyperref}
\usepackage{amsmath}
\usepackage{amssymb}
\definecolor{lightblue}{rgb}{0.66, 0.85, 0.95}
\definecolor{linkcolor}{RGB}{0, 0, 255}
\hypersetup{
    colorlinks=true,
    citecolor=linkcolor,
}
\begin{document}
\title{Opportunistic Osteoporosis Diagnosis via Texture-Preserving Self-Supervision, Mixture of Experts and Multi-Task Integration}

\author{Jiaxing Huang\inst{1,5,\dagger} \and
Heng Guo\inst{2,6,\dagger} \and
Le Lu\inst{2}
\and
Fan Yang\inst{3}
\and
Minfeng Xu\inst{2,6}
\and \\
Ge Yang\inst{1,5,\ddagger}
\and
Wei Luo\inst{4,\ddagger}}
\authorrunning{J. Huang et al.}
%
\institute{Institute of Automation, Chinese Academy of Sciences, Beijing 100190, China \and
DAMO Academy, Alibaba Group\and
Union Hospital, Huazhong University of Science and Technology, China\and
Xiangya Hospital, Central South University, China\and
School of Artificial Intelligence, University of Chinese Academy of Sciences, China\and
Hupan Lab, Hangzhou 310023, China \\
\email{ge.yang@ia.ac.cn, luowei0928@126.com}}

\renewcommand{\thefootnote}{}
\footnotetext{$^\dagger$ Equal Contribution, $^\ddagger$ Corresponding Author.\\
}

\maketitle              

\begin{abstract}
Osteoporosis, characterized by reduced bone mineral density (BMD) and compromised bone microstructure, increases fracture risk in aging populations. While dual-energy X-ray absorptiometry (DXA) is the clinical standard for BMD assessment, its limited accessibility hinders diagnosis in resource-limited regions. Opportunistic computed tomography (CT) analysis has emerged as a promising alternative for osteoporosis diagnosis using existing imaging data. Current approaches, however, face three limitations: (1) underutilization of unlabeled vertebral data, (2) systematic bias from device-specific DXA discrepancies, and (3) insufficient integration of clinical knowledge such as spatial BMD distribution patterns. To address these, we propose a unified deep learning framework with three innovations. First, a self-supervised learning method using radiomic representations to leverage unlabeled CT data and preserve bone texture. Second, a Mixture of Experts (MoE) architecture with learned gating mechanisms to enhance cross-device adaptability. Third, a multi-task learning framework integrating osteoporosis diagnosis, BMD regression, and vertebra location prediction. Validated across three clinical sites and an external hospital, our approach demonstrates superior generalizability and accuracy over existing methods for opportunistic osteoporosis screening and diagnosis.


\end{abstract}

\section{Introduction}
Osteoporosis, marked by reduced bone mineral density (BMD) and deteriorated bone microstructure, predisposes affected individuals to increased fracture susceptibility \cite{on2001osteoporosis}. With global aging populations, both the prevalence of osteoporosis and fracture-related socioeconomic burdens are projected to ascend substantially \cite{burge2007incidence}. Current clinical guidelines for fracture risk stratification integrate quantitative BMD assessment via dual-energy X-ray absorptiometry (DXA) with established clinical risk parameters \cite{jang2019opportunistic,haseltine2021bone}. However, significant diagnostic disparities persist in developing countries due to limited DXA accessibility \cite{siris2001identification,haseltine2021bone}. This disparity underscores the urgency to develop accessible alternatives using existing imaging modalities. Notably, opportunistic computed tomography (CT) analysis, leveraging existing imaging data acquired for alternative clinical indications, has emerged as a promising strategy for osteoporosis diagnosis without requiring additional radiation exposure \cite{pickhardt2013opportunistic,paderno2024artificial}.

Previous studies have demonstrated the feasibility of opportunistic osteoporosis screening and BMD quantification on clinically acquired CT imaging wherein the lumbar vertebrae are key regions suitable for diagnosis of osteoporosis \cite{krishnaraj2019simulating,tariq2023opportunistic,chen2023automatic}. Existing methodologies exhibit three critical limitations that impede real-world clinical generalization. First, previous approaches do not adequately leverage unlabeled images. In Figure \ref{fig1}(A), a subset of vertebrae in CT scans have the corresponding DXA measurements, typically L1-L4 \cite{el2008dxa}. Spine imaging data from chest and abdominal CT scans are abundant in hospital databases, but most unlabeled vertebral CT data remain underutilized \cite{yang2022opportunistic}. Moreover, since osteoporosis manifests itself through changes in bone texture, there is an urgent need for a texture-preserving self-supervised learning algorithm to fully harvest the unlabeled vertebral dataset. Second, systematic bias arises from device-specific discrepancies in DXA measurements. Inter-manufacturer variability among DXA devices can yield divergent results for the same patients \cite{genant1994universal}, and these discrepancies persist despite standardization efforts \cite{fan2010does} as demonstrated in Figure \ref{fig1}(A). Current AI models for simulating DXA measurements from CT scans mostly focus on single-device scenarios, neglecting the inherent heterogeneity and complexity of practical environments. Consequently, models trained on single-device DXA labeled data exhibit compromised generalizability in multi-device clinical settings. Third, conventional architectures do not incorporate the clinical priors essential for accurate osteoporosis diagnosis. Notably, positional variations in BMD distribution across vertebrae~\cite{slosman1990vertebral} – a critical diagnostic feature – are not yet incorporated into existing models. These limitations underscore the need for more robust methodologies to improve real-world applicability.

To overcome these limitations, we propose a unified deep learning framework with three novel contributions. First, we develop a texture-preserving self-supervised learning method based on SwAV \cite{caron2020unsupervised}. Unlike conventional random cropping operations that may compromise structural integrity, our approach uses a sliding window mechanism to extract radiomic features from the original CT images. This modification effectively employs unlabeled data for model pre-training. Second, we develop a Mixture of Experts (MoE) \cite{yuksel2012twenty} architecture in the decoding part to enhance the model's adaptability across multiple devices. This design enables dynamic selection of predictors via learned gating mechanisms. Third, given the clinical importance of spatial bone mineral distribution patterns and quantitative BMD values in osteoporosis assessment, a multi-task learning loss is used, synergistically integrating differential diagnosis, BMD regression and anatomical location prediction. This tripartite architecture improves the precision of osteoporosis diagnosis through complementary supervision. The proposed framework is validated in three independent clinical sites and an external hospital, demonstrating superior generalizability compared to existing methods.

\section{Method}
\subsection{Texture-preserving Self-supervised Learning Framework}

\begin{figure}[t]
\includegraphics[width=\textwidth]{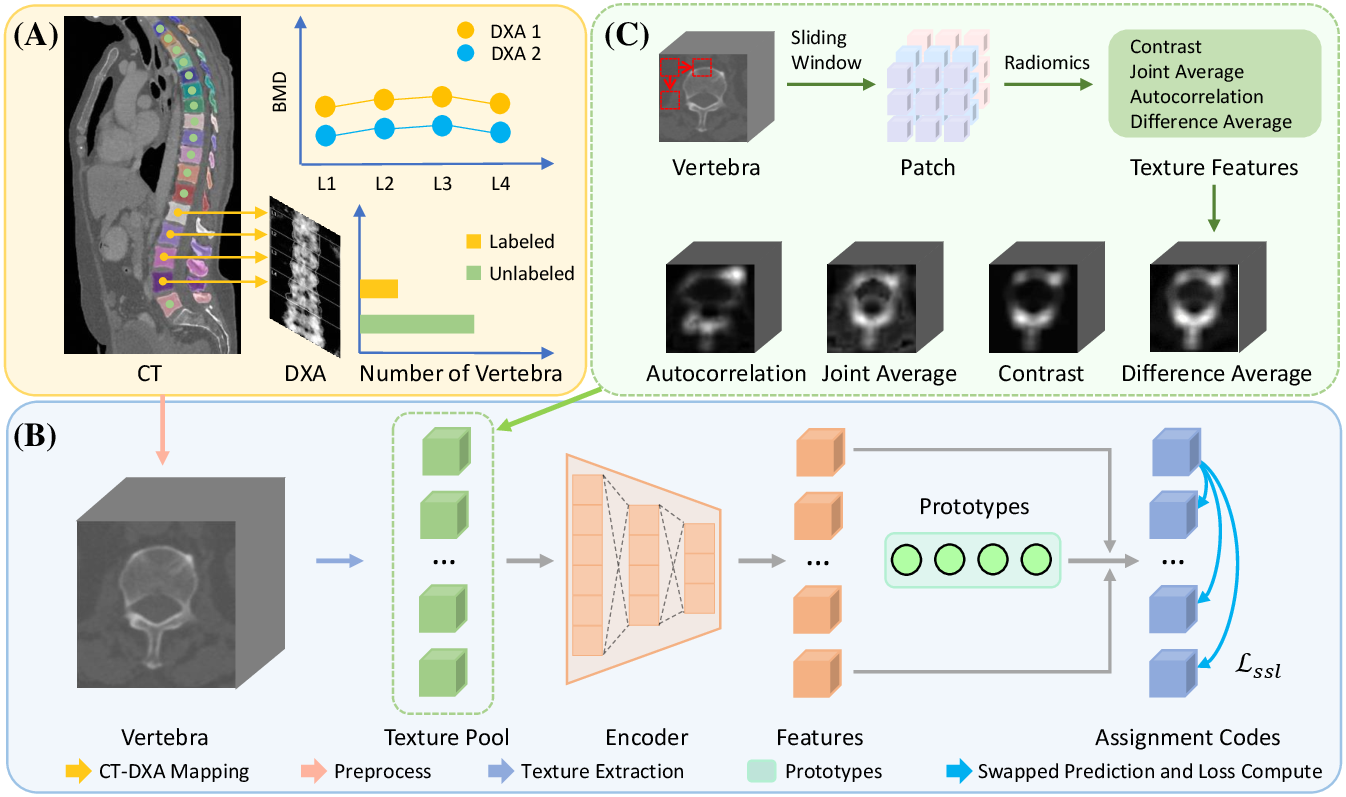}
\caption{(A) Data Acquisition: A localization model identified lumbar regions on CT scans corresponding to paired DXA measurements, correlating vertebra-level CT and DXA data. Notably, different DXA devices may produce varying BMD values for the same vertebra of a patient. (B) Overview of the texture-preserving self-supervised learning framework. (C) Workflow for constructing the texture feature pool.} \label{fig1} 
\end{figure}

Self-supervised learning (SSL) has shown great potential in medical image analysis to leverage unlabeled data~\cite{wang2023swinmm,jiang2023anatomical,wu2024voco}. However, its application to osteoporosis diagnosis faces a critical challenge: texture features, which are pivotal for characterizing bone mineral density patterns, may be compromised by conventional random cropping operations in standard SSL frameworks. To this end, we propose a texture-preserving SSL method (TP-SwAV), as in Figure \ref{fig1}(B). Our approach consists of three key components. Firstly, we establish a texture pool containing both the original image and its texture features including autocorrelation, joint average, contrast, and difference average. We extract these spatially localized texture descriptors \{${\mathbf{TF}}_{1}$,...,${\mathbf{TF}}_{M}$\} through a sliding window mechanism with adaptive interpolation \cite{huang2024representing} as shown in Figure \ref{fig1}(C) via the Pyradiomics toolkit \cite{van2017computational}. This ensures dimensional consistency between the radiomic features and the original image. Secondly, the texture features \{${\mathbf{TF}}_{1}$,...,${\mathbf{TF}}_{M}$\} are encoded into latent embeddings \{${\mathbf{P}}_{1}$,...,${\mathbf{P}}_{M}$\} through a general encoder.  These embeddings will be assigned to a set of $K$ learnable prototypes \{${\mathbf{C}}_{1}$,...,${\mathbf{C}}_{K}$\}. For example, for ${\mathbf{P}}_{t}$ and ${\mathbf{P}}_{s}$, generated by two different views of an image, we compute their prototype assignment distributions ${\mathbf{Q}}_{t}$ and ${\mathbf{Q}}_{s}$ via optimal transport matching \cite{cuturi2013sinkhorn,caron2020unsupervised}. Finally, a swapped prediction loss enforces cross-view consistency:

\begin{equation}
\mathcal{L}_{ssl} = -\sum_{k} \mathbf{Q}_{s}^{(k)} \log \frac{\exp{(\frac{1}{\tau}\mathbf{P}_{t}^{\top}\mathbf{C}_k)}}{\sum_{k^{'}} \exp(\frac{1}{\tau}\mathbf{P}_{t}^{\top}\mathbf{C}_{k^{'}})} - \sum_{k}\mathbf{Q}_{t}^{(k)} \log \frac{\exp{(\frac{1}{\tau}\mathbf{P}_{s}^{\top}\mathbf{C}_k)}}{\sum_{k^{'}} \exp(\frac{1}{\tau}\mathbf{P}_{s}^{\top}\mathbf{C}_{k^{'}})}
\end{equation}
where $\tau$ is a temperature parameter. Simultaneously, a clustering regularization term ensures prototype diversity.

\subsection{Mixture of Experts Architecture for Cross-Device Adaptability}
\begin{figure}[t]
\includegraphics[width=\textwidth]{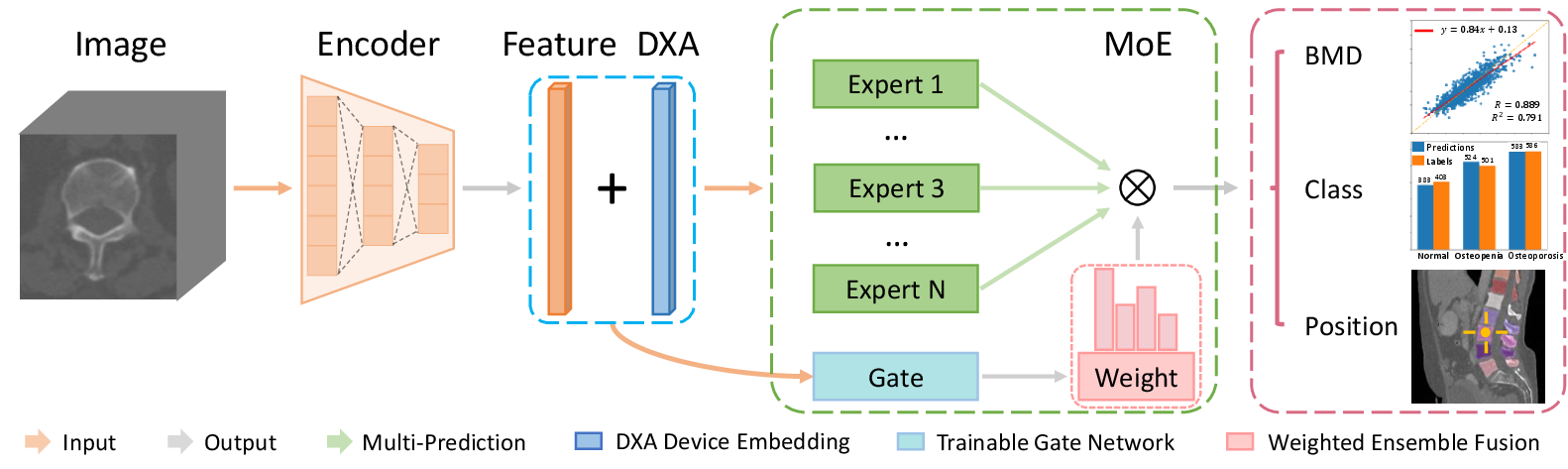}
\caption{Overflow of MoE-enhanced multi-task learning framework.} \label{fig2} 
\end{figure}

To address the inherent heterogeneity across DXA devices, we propose a Mixture of Experts (MoE) architecture integrated into the decoding stage. As shown in Figure \ref{fig2}, the framework comprises two core components: (1) Device-specific expert modules: A set of $N$ ($N=3$) specialized expert heads \{${E}_{1}$,...,${E}_{N}$\} that capture device-specific features. (2) Adaptive fusion mechanism: A dynamic gating network that generates adaptive attention weights based on the sum of image embeddings $\mathbf{P}_{I}$ and the corresponding DXA device embeddings $\mathbf{D}_I$ of image I. The gating operation is formulated as:
\begin{equation}
\alpha_i = \mathrm{Softmax}\big(f_g(\mathbf{P}_I + \mathbf{D}_I)\big),\quad \sum_{i=1}^N \alpha_i = 1
\end{equation}
where ${f}_{g}(\cdot)$ denotes the parameterized gating function. The final prediction synthesizes expert outputs through weighted summation:
\begin{equation}
\mathrm{Output} = \sum_{i=1}^N \alpha_i \cdot E_i
\end{equation}
Introducing MoE structure at the decoding stage provides two key benefits to the model: (1) \textbf{Adaptive Feature Fusion:} The gating network automatically activates relevant experts via attention weights derived from input image embeddings and corresponding DXA device embeddings, enabling dynamic adaptation to different devices. (2) \textbf{Cross-Device Regularization:} Joint optimization of both expert-specific parameters and shared gating parameters leads to an implicit regularization effect, which prevents the model overfitting to any single device while maintaining discriminative power.

\subsection{Multi-task Learning with Complementary Supervision}
Our fine-tuning framework integrates three complementary supervision signals through multi-task learning, as illustrated in Figure \ref{fig2}. First, a BMD regression branch directly predicts continuous bone mineral density values from encoded image features, ensuring the model captures biologically meaningful patterns critical for osteoporosis assessment. Second, a classification branch models vertebral anatomical landmarks (T1-T12, L1-L6) as ordinal labels (1-18) to induce position-related regularization, thereby enabling the model to perceive vertebral positions, as vertebra at different positions exhibit distinct BMD distributions. Third, the primary diagnostic task employs cross-entropy loss for osteoporosis detection, enhanced by shared representations from auxiliary tasks. The unified objective function combines these components:
\begin{equation}
\mathcal{L}_{unified} = \mathcal{\alpha}\mathcal{L}_{class}  + \mathcal{\beta}\mathcal{L}_{BMD} +\mathcal{\gamma}\mathcal{L}_{position}
\end{equation}
where $\mathcal{L}_{class}$ and $\mathcal{L}_{position}$ both utilize CE loss and $\mathcal{L}_{BMD}$ employs MSE loss. Hyperparameters ($\alpha=1.0$, $\beta=1.0$, $\gamma=0.5$) balance task-specific gradients, enabling synergistic learning of diagnostic features through shared weights.

\begin{table}[t]
  \centering
  \caption{Dataset composition across four clinical sites. D* represents the external site.}
  \resizebox{0.8\textwidth}{!}{
  \small
    \begin{tabular}{c|c|ccc|c|ccc}
     \Xhline{1pt}
      \multirow{2}[1]{*}{\textbf{Site}} & \multirow{2}[1]{*}{\textbf{DXA Devices}} & \multicolumn{3}{c|}{\textbf{Patient}} & \multicolumn{1}{c|}{\textbf{Vertebra w/o GT}}& \multicolumn{3}{c}{\textbf{Vertebra w/ GT}}  \\
   \cline{3-9} 
    &  & Train & Valid & Test  & Pre-train  & Train & Valid & Test \\
    \hline
    A     & Hologic & 1330  & 226  & 669 & 18308 & 2980  & 510   & 1490 \\
    B     & GE Lunar & 683   & 112   & 325 & 9232 & 1220  & 200   & 610 \\
    C     & MEDIX & 217   & 36    & 106  & 2716& 801   & 129   & 398 \\
    D* & Hologic & 0     & 0     & 193  & 0& 0     & 0     & 271 \\
    ALL   & Mix     & 2230  & 374   & 1293 & 30256& 5001  & 839   & 2769 \\
    \Xhline{1pt}
    \end{tabular}}
  \label{table1} 
\end{table}%

\begin{table}[h]
  \centering
  \caption{Vertebra-level quantitative results of osteoporosis diagnosis.}
  \resizebox{0.9\textwidth}{!}{
 \begin{tabular}{c|c|c|ccccc|ccccc} 
 \Xhline{1pt}
    \multirow{2}[0]{*}{Model} & \multirow{2}[0]{*}{Pre-train} & \multirow{2}[0]{*}{Loss} & \multicolumn{5}{c|}{Site A}   & \multicolumn{5}{c}{Site B} \\
   \cline{4-13} 
          &       &       & ACC   & SEN   & SPE   & F1 & P   & ACC   & SEN   & SPE   & F1& P \\
          \hline
    ResNet34 & /     & $\mathcal{L}_{class} $   & 83.89 & 76.86 & 89.24 & 80.49 & *** & 86.56 & 62.24 & 91.21 & 59.80& *** \\
    Swin-UNTER & /     & $\mathcal{L}_{class} $   & 84.70  & \cellcolor{pink!40}83.52& 85.36 & 79.64 & *** & \cellcolor{pink!40}88.69 & \cellcolor{lightblue}\pmb{76.06} & 90.35 & 61.02 &*** \\
    ViT   & MAE   & $\mathcal{L}_{class} $   & 81.34 & 70.21 & \cellcolor{lightblue}\pmb{92.99} & 79.38&*** & 82.62 & 50.00    & \cellcolor{pink!40}92.64& 59.23 &* \\
    nnFormer & Alice & $\mathcal{L}_{class} $   & \cellcolor{pink!40}85.50  & 79.00    & \cellcolor{pink!40}90.38 & \cellcolor{pink!40}82.35&***& 88.52 & 
 \cellcolor{pink!40}70.93& 91.41 & 63.54&*** \\
    SwinT & SwinMM & $\mathcal{L}_{class} $   & 85.37 & \cellcolor{lightblue}\pmb{86.36} & 84.86 & 80.04 &*** & 87.38 & 63.81 & 92.28 & 63.51&*** \\
    ResNet34 & VOCO  & $\mathcal{L}_{class} $   & 83.83 & 79.39 & 86.71 & 79.45 &*** & 87.21 & 64.89 & 91.28 & 61.00 &***\\
    ResNet34 & SwAV  & $\mathcal{L}_{class} $   & 83.36 & 78.84 & 86.28 & 78.84 &*** & 87.05 & 66.67 & 90.17 & 57.75 &***\\
    ResNet34 & TP-SwAV & $\mathcal{L}_{class} $   & 84.43 & 80.94 & 86.60  & 79.97 & / & 86.72 & 62.89 & 91.23 & 60.10 &/ \\
    ResNet34 & TP-SwAV & $\mathcal{L}_{unified} $  & 84.76 & 81.44 & 86.83 & 80.38 &/ & 88.19 & 67.00    & 92.35 &\cellcolor{pink!40}65.05 &/\\
    ResNet34+MoE & TP-SwAV & $\mathcal{L}_{unified}$   & \cellcolor{lightblue}\pmb{86.64}& 83.19 & 88.86 &  \cellcolor{lightblue}\pmb{82.98} & Base & \cellcolor{lightblue}\pmb{89.02}& 69.31 & \cellcolor{lightblue}\pmb{92.92} & \cellcolor{lightblue}\pmb{67.63}  & Base\\
    \hline
    \multirow{2}[0]{*}{Model} & \multirow{2}[0]{*}{Pre-train} & \multirow{2}[0]{*}{Loss} & \multicolumn{5}{c|}{Site C}   & \multicolumn{5}{c}{Site D} \\\cline{4-13} 
          &       &       & ACC   & SEN   & SPE   & F1 & P   & ACC   & SEN   & SPE  & F1 & P \\
          \hline
    ResNet34 & /     & $\mathcal{L}_{class} $   & 89.20  & 71.43 & 92.11 & 65.04 & * & 73.80  & 78.70  & 70.55 & 70.54 &*** \\
    Swin-UNTER & /     & $\mathcal{L}_{class} $   & 89.95 & 81.40  & 90.99 & 63.64 &*** & 73.43 & 92.96 & 66.50  & 64.71  &*** \\
    ViT   & MAE   & $\mathcal{L}_{class} $   & 87.18 & 59.09 &\cellcolor{pink!40}93.86& 67.10 &*** & 78.59 & 76.22 & \cellcolor{lightblue}\pmb{81.25} &\cellcolor{pink!40}78.99 &**\\
    nnFormer & Alice & $\mathcal{L}_{class} $   & \cellcolor{lightblue}\pmb{91.21} & \cellcolor{lightblue}\pmb{88.10} & 91.57 & 67.89&*** & 79.70  & 90.62 & 73.71 & 75.98&** \\
    SwinT & SwinMM & $\mathcal{L}_{class} $   & 89.45 & \cellcolor{pink!40}87.88& 89.59 & 58.00  &**  & \cellcolor{pink!40}80.07& 92.47 & 73.60  & 76.11&*** \\
    ResNet34 & VOCO  & $\mathcal{L}_{class} $   & 88.44 & 69.81 & 91.30  & 61.67 &* & 74.54 & 81.37 & 70.41 & 70.64 &**\\
    ResNet34 & SwAV  & $\mathcal{L}_{class} $   & 88.44 & 68.42 & 91.79 & 62.90 &*  & 79.70  & \cellcolor{pink!40}93.33& 72.93 & 75.34 &* \\
    ResNet34 & TP-SwAV & $\mathcal{L}_{class} $   & 90.70  & 77.78 & 92.73 & 69.42 &/ & 78.97 & \cellcolor{lightblue}\pmb{94.19} & 71.89 & 73.97 &/ \\
    ResNet34 & TP-SwAV & $\mathcal{L}_{unified}$   & \cellcolor{lightblue}\pmb{91.21} & 76.67 & 93.78 & \cellcolor{pink!40}72.24 & / & 79.33 & 87.83 & 74.40  & 76.27 &/\\
    ResNet34+MoE & TP-SwAV & $\mathcal{L}_{unified}$   & \cellcolor{pink!40}90.95& 74.60  & \cellcolor{lightblue}\pmb{94.02} & \cellcolor{lightblue}\pmb{72.31} & Base & \cellcolor{lightblue}\pmb{82.65} & 85.83 & \cellcolor{pink!40}80.13&\cellcolor{lightblue}\pmb{81.42} & Base\\
    \Xhline{1pt}
    \end{tabular}}
  \label{tab2}%
\end{table}%
\section{Experiments}
\subsection{Dataset and Evaluation Metrics}
We conducted a multicenter retrospective study utilizing data from three tertiary hospitals (Sites A-C, March 2018-December 2023) for model development and one external testing site (Site D, April 2022-December 2023). Inclusion criteria required paired non-contrast chest/abdominal CT and dedicated DXA scans within six months. Exclusion criteria included: 1) poor image quality; 2) metal implants; 3) vertebral compression fractures; 4) failed automated vertebrae identification and 5) missing lumbar spine data.

CT scans were obtained using standard non-contrast protocols across multiple vendors (Siemens, GE, Philips, etc.). DXA measurements (L1-L4 vertebrae) were performed using Hologic, GE Lunar, and MEDIX systems by certified technicians following standardized protocols. The Med-Query algorithm \cite{guo2024med} established spatial CT-DXA correlations via automated vertebral localization and rotated ROI extraction, enabling precise anatomical mapping and corresponding BMD quantification. As shown in Table \ref{table1}, the final dataset comprised 2230 patients (5001 vertebrae) for training, 374 (839 vertebrae) for validation, 1100 patients (2498 vertebrae) for internal testing and 193 patients (271 vertebrae) for external testing on Site D. Additionally, 30256 unlabeled vertebrae were used for pre-training. For comparative analysis, we selected two classification models trained from scratch (ResNet34 \cite{he2016deep} and Swin-UNETR \cite{hatamizadeh2021swin}) as baseline methods, along with several representative self-supervised learning (SSL) approaches. These include two general-purpose SSL methods (MAE \cite{he2022masked} and SwAV \cite{caron2020unsupervised}) and three recently proposed medical imaging-oriented methods (SwinMM \cite{wang2023swinmm}, Alice \cite{jiang2023anatomical}, and VOCO \cite{wu2024voco}). Due to architectural constraints of certain self-supervised frameworks, the backbones were not fully standardized across methods. Specifically, nnFormer \cite{zhou2023nnformer} was used in Alice, Vision Transformer (ViT) \cite{ranftl2021vision} in MAE, and Swin Transformer (SwinT) \cite{liu2021swin} in SwinMM.

All models were trained under standardized hyperparameters to ensure comparability. For self-supervised pre-training, configurations included: batch size=32, AdamW optimizer (${\beta}_{1}=0.9$, ${\beta}_{2}=0.999$, weight decay=0.1), and linear warmup learning rate scheduling (initial lr=1e-4, warmup epochs=100) over 300 total epochs. During fine-tuning, we maintained identical batch-size and optimizer parameters while adjusting learning dynamics: reduced initial learning rate (lr=1e-5), extended warmup duration (150 epochs), and increased the maximum training cycles to 400 epochs. Classification performance was quantified using four metrics: accuracy (ACC), sensitivity (SEN), specificity (SPE), and F1-score (F1). In this study, the evaluation metrics ACC, SEN, SPE, and F1, are expressed as percentages (\%). Patient-level diagnostic results were derived from vertebra-level predictions. Specifically, for a patient with $N$ vertebrae, the patient was classified as osteoporosis if any one of the vertebrae was diagnosed or predicted as osteoporosis. This principle guided the computation of patient-level outcomes.

\begin{table}[t]
  \centering
  \caption{Patient-level quantitative results of osteoporosis diagnosis.}
  \resizebox{0.8\textwidth}{!}{
    \begin{tabular}{c|c|c|cccc|cccc}
    \Xhline{1pt}
    \multirow{2}[0]{*}{Model} & \multirow{2}[0]{*}{Pre-train} & \multirow{2}[0]{*}{Loss} & \multicolumn{4}{c|}{Site A}   & \multicolumn{4}{c}{Site B} \\
    \cline{4-11} 
          &       &       & ACC   & SEN   & SPE   & F1    & ACC   & SEN   & SPE   & F1 \\
          \hline
    ResNet34 &    /   & $\mathcal{L}_{class} $   & 83.89 & 77.18 & 89.55 & 81.43 & 84.65 & 76.09 & 89.81 & 78.85 \\
    Swin-UNTER &   /    & $\mathcal{L}_{class} $   & 85.44 & \cellcolor{pink!40}85.22 & 85.57 & 81.44 & 85.92 & \cellcolor{pink!40}84.85 & 86.38 & 78.32 \\
    ViT   & MAE   & $\mathcal{L}_{class} $   & 83.86 & 76.79 & \cellcolor{pink!40}90.99& 82.69 & 81.85 & 75.55 & 90.61 & 69.86 \\
    nnFormer & Alice & $\mathcal{L}_{class} $   & \cellcolor{pink!40}86.08 & 82.61 & \textit{88.55} & \cellcolor{pink!40}83.12 & 86.38 & 82.02 & 88.56 & 80.05 \\
    SwinT & SwinMM & $\mathcal{L}_{class} $   & 85.80  & \cellcolor{lightblue}\pmb{89.39} & 83.73 & 82.18 & 87.69 & 71.43 & 91.60  & 69.23 \\
    ResNet34 & VOCO  & $\mathcal{L}_{class} $   & 83.51 & 79.32 & 86.50  & 80.06 & 84.38 & 78.10  & 87.67 & 77.49 \\
    ResNet34 & SwAV  & $\mathcal{L}_{class} $   & 83.76 & 78.92 & 87.39 & 80.62 & 84.38 & 77.66 & 87.99 & 77.66 \\
    ResNet34 & TP-SwAV & $\mathcal{L}_{class} $   & 85.82 & 82.91 & 87.83 & 82.65 & 86.01 & 80.80  & 88.71 & 79.74 \\
    ResNet34 & TP-SwAV & $\mathcal{L}_{unified}$   & 84.60  & 79.94 & 88.61 & 82.75 & \cellcolor{pink!40}92.00    & 81.54 & \cellcolor{pink!40}94.62 & \cellcolor{pink!40}80.30 \\
 ResNet34+MoE & TP-SwAV & $\mathcal{L}_{unified}$   & \cellcolor{lightblue}\pmb{87.74} & 83.10 & \cellcolor{lightblue}\pmb{91.88} & \cellcolor{lightblue}\pmb{86.33} & \cellcolor{lightblue}\pmb{93.23} & \cellcolor{lightblue}\pmb{86.89} & \cellcolor{lightblue}\pmb{94.70} & \cellcolor{lightblue}\pmb{82.81} \\
 \hline
  \multirow{2}[1]{*}{Model} & \multirow{2}[1]{*}{Pre-train} & \multirow{2}[1]{*}{Loss} & \multicolumn{4}{c|}{Site C}   & \multicolumn{4}{c}{Site D} \\ \cline{4-11} 
          &       &       & ACC   & SEN   & SPE   & F1    & ACC   & SEN   & SPE   & F1 \\ \hline
    ResNet34 &    /   & $\mathcal{L}_{class} $   & 84.91 & 74.07 & 88.61 & 71.43 & 82.93 & 76.00    & \cellcolor{lightblue}\pmb{87.30}  & 77.47 \\
    Swin-UNTER &   /    & $\mathcal{L}_{class} $   & 83.02 & 78.95 & 83.91 & 62.50  & 83.55 & 85.45 & 82.77 & 75.20 \\
    ViT   & MAE   & $\mathcal{L}_{class} $   & 83.96 & 70.00    & 89.47 & 71.19 & 79.79 & 82.56 & 77.57 & 78.48 \\
    nnFormer & Alice & $\mathcal{L}_{class} $   & 86.79 & \cellcolor{lightblue}\pmb{94.12} & 85.39 & 69.57 & \cellcolor{lightblue}\pmb{85.10}  & 83.49 & \cellcolor{pink!40}85.90  & 78.81 \\
    SwinT & SwinMM & $\mathcal{L}_{class} $   & 81.13 & \cellcolor{pink!40}90.91 & 80.00    & 67.80  & 79.79 & \cellcolor{lightblue}\pmb{91.18}& 73.60 & 76.07 \\
    ResNet34 & VOCO  & $\mathcal{L}_{class} $   & 79.25 & 78.12 & 83.13 & 67.69 & 83.24 & 79.20  & 85.41 & 76.74 \\
    ResNet34 & SwAV  & $\mathcal{L}_{class} $   & 86.79 & 80.00    & 88.89 & 74.07 & 83.40  & 79.56 & 85.44 & 76.91 \\
    ResNet34 & TP-SwAV & $\mathcal{L}_{class} $   & \cellcolor{pink!40}86.82 & 80.10  & 89.18 & 74.27 & 84.63 & \textit{82.34} & 85.80  & 78.30 \\
    ResNet34 & TP-SwAV & $\mathcal{L}_{unified}$   & \cellcolor{lightblue}\pmb{88.68} & 84.00    & \cellcolor{pink!40}90.12 &\cellcolor{pink!40}77.78& 81.87 & 88.46 & 77.39 & \cellcolor{pink!40}79.77 \\
 ResNet34+MoE & TP-SwAV & $\mathcal{L}_{unified}$   & \cellcolor{lightblue}\pmb{88.68}& 81.48 & \cellcolor{lightblue}\pmb{91.14} & \cellcolor{lightblue}\pmb{78.57} & \cellcolor{pink!40}84.97 & \cellcolor{pink!40}90.24 & 81.08 & \cellcolor{lightblue}\pmb{83.62} \\
    \Xhline{1pt}
    \end{tabular}%
    }
  \label{tab3}%
\end{table}%

\subsection{Main Results}
\begin{figure}[t]
\includegraphics[width=\textwidth]{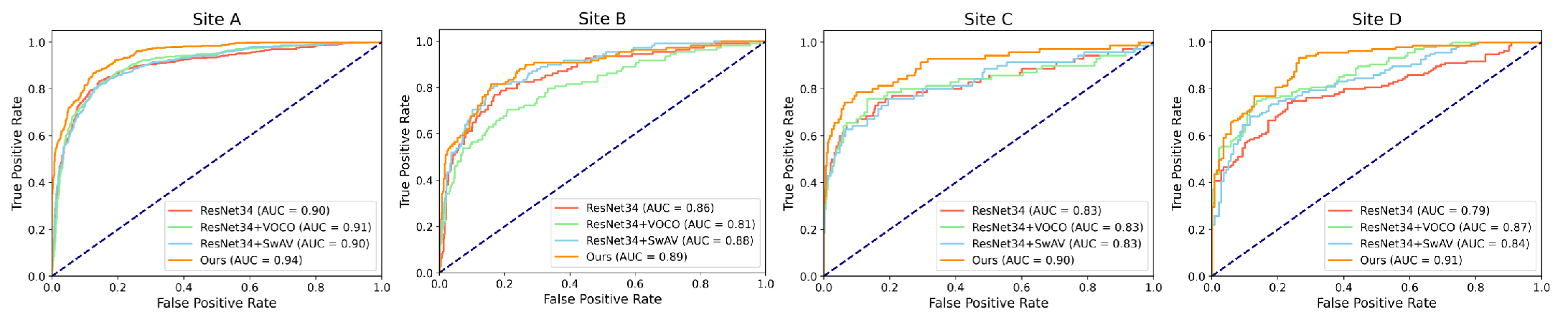}
\caption{Receiver operating characteristic (ROC) curves.} \label{fig3} 
\end{figure}

The experimental results presented in Table~\ref{tab2} and Table~\ref{tab3} lead to three principal findings: (1) Compared to SwAV, the proposed TP-SwAV substitutes conventional cropping operations with pre-computed feature pool extraction. This modification significantly enhances diagnostic performance across all four clinical centers, substantiating the critical role of texture information in osteoporosis assessment. (2) Deep supervision through concurrent bone mineral density regression and anatomical position prediction demonstrates substantial improvements in diagnostic accuracy, confirming the synergistic benefits of multi-task learning frameworks. (3) The Mixture-of-Experts (MoE) architecture demonstrates superior adaptability to multi-center/multi-device scenarios. When integrated with multi-task learning and TP-SwAV, the ResNet34-based architecture achieves state-of-the-art F1-scores in both vertebra-level and patient-level osteoporosis diagnoses. Notably, patient-level validation yields F1-scores of $86.33\%$ (Site A), $82.81\%$ (Site B), $78.57\%$ (Site C) for internal cohorts, and $83.62\%$ for external testing (Site D). (4) We performed paired t-tests to compare our method with other approaches. The results presented in Table \ref{tab2} demonstrate statistically significant differences compared to these methods across all four sites. The significance levels are denoted as follows: `*' indicates $p<0.05$, `**' represents $p<0.01$, and `***' corresponds to $p<0.001$. In addition, we constructed receiver operating characteristic (ROC) curves for the ResNet34-based methods to provide additional validation of our approach's efficacy as shown in Figure \ref{fig3}.

\begin{table}[t]
  \centering
  \begin{minipage}{0.5\textwidth}
    \centering
    \caption{Ablation study of loss function.}
    \resizebox{0.92\textwidth}{!}{%
      \begin{tabular}{c|c|cccc}
        \Xhline{1pt}
        Model & Loss  & Site A & Site B & Site C & SiteD \\
        \hline
        ResNet34 & $\mathcal{L}_{class} $   & 80.49 & 59.80  & 65.04 & 70.54 \\
        ResNet34 & $\mathcal{L}_{class}+\mathcal{L}_{BMD} $ & 80.87 & 60.64 & 68.47 & 70.74 \\
        ResNet34 & $\mathcal{L}_{class}+\mathcal{L}_{position} $ & 80.54 & 61.49 & 65.62 & 71.52 \\
        ResNet34 & $\mathcal{L}_{unified}$ & \pmb{81.18} & \pmb{61.70}  & \pmb{70.97} & \pmb{72.18} \\
        \Xhline{1pt}
      \end{tabular}%
    }
    \label{tab4}%
  \end{minipage}\hfill
  \begin{minipage}{0.5\textwidth}
    \centering
    \caption{Ablation study of backbone.}
    \resizebox{\textwidth}{!}{%
      \begin{tabular}{c|c|c|cccc}
        \Xhline{1pt}
        Model  & Pre-train & Loss  & Site A & Site B & Site C & SiteD \\
        \hline
        SwinT & /     & $\mathcal{L}_{class} $    &  79.14     &   60.69 &     56.46   & 70.19 \\
        SwinT & SwinMM & $\mathcal{L}_{class} $    & 80.04 & 63.51 & 58.00    & 76.11 \\
        SwinT & TP-SwAV & $\mathcal{L}_{class} $    & 80.00    & 61.05 & 58.33 & \pmb{82.44} \\
        SwinT+MoE & TP-SwAV & $\mathcal{L}_{unified} $  & \pmb{80.98} & \pmb{64.00}    & \pmb{60.18} & 80.31 \\
        \Xhline{1pt}
      \end{tabular}%
    }
    \label{tab5}%
  \end{minipage}
\end{table}

\subsection{Ablation Study}
To validate the efficacy of multi-task supervision, we conducted experiments using ResNet34 (Table \ref{tab4}), demonstrating that both BMD regression ($\mathcal{L}_{BMD}$) and position prediction ($\mathcal{L}_{position}$) significantly improved performance, as indicated by increased F1-scores over the single-task baseline. Furthermore, to assess generalizability, we implemented our framework with Swin Transformer as the backbone. Results in Table \ref{tab5} show consistent performance gains (F1-score) across four clinical datasets when incorporating TP-SwAV, MoE structure, and the proposed unified loss ($\mathcal{L}_{unified}$), confirming the adaptability of our approach to both CNN and Transformer architectures.

\section{Conclusion}
This study presents a unified deep learning framework addressing three critical challenges in opportunistic osteoporosis diagnosis using CT imaging. First, the texture-preserving TP-SwAV algorithm effectively utilizes unlabeled vertebral data through extracted radiomics features, overcoming limitations of random cropping operations in conventional SSL methods. Second, the MoE architecture with adaptive gating mechanisms mitigates device-specific variability in DXA measurements, enhancing cross-device generalizability. Third, multi-task learning integrating BMD regression, position prediction, and diagnostic classification leverages complementary clinical priors to improve diagnostic accuracy. Validated across multiple centers, our framework demonstrates superior adaptability and robustness, providing a scalable solution for osteoporosis screening in resource-constrained regions and reducing reliance on DXA equipment.

\begin{credits}
\subsubsection{\ackname} This work was supported by the National Natural Science Foundation of China (grant 92354307), the National Key R\&D Program of China (grant 2024YFF0729202), the Strategic Priority Research Program of the CAS (XDA0460305), and the Fundamental Research Funds for the Central Universities (grant E3E45201X2). Support was also provided by Alibaba Group via the Alibaba Research Intern Program.
\subsubsection{\discintname}
The authors have no competing interests to declare that are relevant to the content of this article.
\end{credits}

%
%
%
\bibliographystyle{splncs04}
\bibliography{main}
%

\end{document}